\documentstyle[12pt]{article}
\def\nab#1{{\nabla_{#1}}}
\def\nabstar#1{\nabla\kern-0.5pt\smash{\raise 4.5pt\hbox{$\ast$}}
               \kern-4.5pt_{#1}}

\def\drvstar#1{\partial\kern-0.5pt\smash{\raise 4.5pt\hbox{$\ast$}}
               \kern-5.0pt_{#1}}

\def\newline{\relax\ifhmode\null\hfil\break\else\nonhmodeerr@\newline\fi}
\def\frac#1#2{{#1\over#2}}
\def\text#1{{\hbox{\rm #1}}}
\def\flushpar{{\par \noindent}}

\newcommand{\beq}{\begin{equation}}
\newcommand{\eeq}{\end{equation}}
\newcommand{\bea}{\begin{eqnarray}}
\newcommand{\eea}{\end{eqnarray}}
\def\Id{ \mbox{1\hspace{-1.2mm}I} }
\def\BE{\begin{equation}}
\def\EE{\end{equation}}
\def\BA{\begin{eqnarray}}
\def\EA{\end{eqnarray}}
\def\BAN{\begin{eqnarray*}}
\def\EAN{\end{eqnarray*}}

\def\tr{\mbox{tr}}

\def\det{\mbox{det}}

\def\gm5{\gamma^5}

\def\CT{{\cal T}}
\def\anx{{\cal A}_n(x)}

\input epsf.sty
\newdimen\psfigsize
\def\psfigure#1 #2 #3 #4 #5{
    \begin{figure}[tbh]
      \begin{center}
      \vbox{
        \null\vskip-0.2in\hskip#2
        \epsfxsize=#1
        \epsfbox{#4}
        \vskip -0.3in
        \caption {#5 \label{#3}}
        \vskip 0.0 true in plus 0.3 true in
      }
      \end{center}
   \end{figure}
}
\begin{document}
\thispagestyle{empty}
\begin{flushright}
NTUTH-99-100 \\
October 1999
\end{flushright}
\bigskip\bigskip\bigskip
\vskip 2.5truecm
\begin{center}
{\LARGE {A note on the solutions of the
         Ginsparg-Wilson relation}}
\end{center}
\vskip 1.0truecm
\centerline{Ting-Wai Chiu}
\vskip5mm
\centerline{Department of Physics, National Taiwan University}
\centerline{Taipei, Taiwan 106, Republic of China.}
\centerline{\it E-mail : twchiu@phys.ntu.edu.tw}
\vskip 2cm
\bigskip \nopagebreak \begin{abstract}
\noindent

The role of $R$ in the solutions of the Ginsparg-Wilson relation is
discussed.

\vskip 2cm
\noindent PACS numbers: 11.15.Ha, 11.30.Rd, 11.30.Fs

\end{abstract}
\vskip 1.5cm

\newpage\setcounter{page}1

In this note, we would like to clarify some seemingly subtle
issues pertaining to the role of $R$ in the solutions of
the Ginsparg-Wilson relation \cite{gwr}
\bea
\label{eq:gwr}
D \gamma_5 + \gamma_5 D = 2 D \gamma_5 R D \ .
\eea
Here $ R $ is a positive definite hermitian operator which
is local in the position space and trivial in the Dirac space.
Since we can sandwich (\ref{eq:gwr}) by a left multiplier $ \sqrt{R} $
and a right multiplier $ \sqrt{R} $ on both sides of (\ref{eq:gwr}),
and define $ D' = \sqrt{R} D \sqrt{R} $, then $ D' $
satisfies
\bea
\label{eq:gwr_1}
D' \gamma_5 + \gamma_5 D' = 2 D' \gamma_5 D'
\eea
which is in the same form of Eq. (\ref{eq:gwr}) with $ R = \Id $.
This seems to suggest that one can set $ R = \Id $ in the GW relation
(\ref{eq:gwr}) and completely ignore the $ R $ dependence in the general
solution of the GW relation. However, as we will see, if $ R $ is set to
be the identity operator from the beginning, then some
of the salient features of the general solution may be easily overlooked.

First, let us review some basics of the GW relation.
The general solution to the GW relation (\ref{eq:gwr}) can be written
as \cite{twc98:6a, twc98:9a}
\bea
\label{eq:gwr_sol}
D = D_c ( \Id + R D_c )^{-1} = ( \Id + D_c R )^{-1} D_c
\eea
where $ D_c $ is any chirally symmetric Dirac operator, i.e.,
\bea
\label{eq:Dc}
D_c \gamma_5 + \gamma_5 D_c = 0 \ .
\eea
In order to have $ D $ reproduce the continuum physics, $ D_c $ is required
to satisfy the necessary physical constraints \cite{twc98:9a}.
The general solution of $ D_c $ has been investigated in ref. \cite{twc99:8}.
Conversely, for any $D$ satisfying the GW relation (\ref{eq:gwr}),
there exists the chirally symmetric $ D_c $
\bea
\label{eq:gwr_sol_Dc}
D_c = D ( \Id - R D )^{-1} = ( \Id - D R )^{-1} D \ .
\eea
We usually require that $ D $ also satisfies
the hermiticity condition\footnote{
This implies that $ \det(D) $ is real and non-negative.}
\bea
\label{hermit_D}
D^{\dagger} = \gamma_5 D \gamma_5 \ .
\eea
Then $ D_c $ also satisfies the hermiticity condition
$ D_c^{\dagger} = \gamma_5 D_c \gamma_5 $, since $ R $ is hermitian and
commutes with $ \gamma_5 $. The hermiticity condition together with the chiral
symmetry of $ D_c $ implies that $ D_c $ is anti-hermitian. Thus there exists
one to one correspondence between $ D_c $ and a unitary opertor $ V $ such
that
\bea
D_c = (\Id + V )(\Id - V )^{-1}, \hspace{4mm}
V = (D_c - \Id)( D_c + \Id)^{-1}.
\label{eq:VDc}
\eea
where $ V $ also satisfies the hermiticity condition
$ V^{\dagger} = \gamma_5 V \gamma_5 $.
Then the general solution (\ref{eq:gwr_sol}) can be written as
\cite{twc98:6a, twc98:9a}
\bea
\label{eq:gwsola}
D = ( \Id + V )[ ( \Id - V ) + R ( \Id + V ) ]^{-1} \\
\label{eq:gwsolb}
  = [ (\Id - V ) + ( \Id + V ) R ]^{-1} ( \Id + V ).
\eea

On the other hand, if one starts from Eq. (\ref{eq:gwr_1}), then its
solution is
\bea
\label{eq:sol_1}
D' = \frac{1}{2} ( \Id + V )
\eea
which agrees with (\ref{eq:gwsola}) with $ R=\Id $. Then using
the relation $ D' = \sqrt{R} D \sqrt{R} $, one obtains
\bea
\label{eq:sol_1a}
D = \frac{1}{\sqrt{2R}} ( \Id + V ) \frac{1}{\sqrt{2R}}
\eea
However, (\ref{eq:sol_1a}) is in contradiction with (\ref{eq:gwsola})
since the $ R $ dependence in (\ref{eq:sol_1a}) can be factored out
completely while that of (\ref{eq:gwsola}) can not.
So, one of them can not be true in general.
If we take the limit $ R \to 0 $,
then (\ref{eq:gwsola}) gives that
$ D \to ( \Id + V ) ( \Id - V )^{-1} = D_c $,
but (\ref{eq:sol_1a}) implies that $ D \to \infty $. Since
$ D_c $ is well defined ( without poles ) in the trivial gauge sector,
it follows that (\ref{eq:sol_1a}) can not be true in general.
The fallacy in (\ref{eq:sol_1a}) is due to the assumption that
$ D'$ is independent of $ R $. From (\ref{eq:gwsola}),
one can derive the following formula
\bea
\label{eq:sol_1b}
D = \frac{1}{\sqrt{2R}} ( \Id + V' ) \frac{1}{\sqrt{2R}}
\eea
where
\bea
\label{eq:V'}
V' = \left[ ( \Id + V ) \sqrt{R} + ( \Id - V ) \frac{1}{\sqrt{R}} \right]^{-1}
     \left[ ( \Id + V ) \sqrt{R} - ( \Id - V ) \frac{1}{\sqrt{R}} \right]  \\
   = \left[ \sqrt{R} ( \Id + V ) - \frac{1}{\sqrt{R}} ( \Id - V ) \right]
     \left[ \sqrt{R} ( \Id + V ) + \frac{1}{\sqrt{R}} ( \Id - V ) \right]^{-1}
\eea
which is unitary and depends on $ R $. This shows that $ D' $ actually
depends on $ R $ and equals to $ \frac{1}{2} ( \Id + V' ) $
rather than (\ref{eq:sol_1}).
Therefore it is {\it erroneous} to write the general solution of the
GW relation in the form of (\ref{eq:sol_1a}), in which $ V' $ is
replaced by $ V $.
Consequently, (\ref{eq:sol_1a}) may mislead one to infer that
$ R $ does not play any significant roles in the locality of $ D $,
in particular when $ R $ is proportional
to the identity operator. However, $ R $ indeed plays a very
important role in determining the locality of $ D $.
Let us consider $ R = r \Id $ with $ r > 0 $.
When $ r \to 0 $, $ D \to D_c $ which must be nonlocal if $ D_c $ is
free of species doubling and has the correct behavior in the classical
continuum limit \cite{no-go}.
As the value of $ r $ moves away from zero and goes towards a finite
value, $ D $ may change from a non-local operator to a local operator.
This has been demonstrated in ref. \cite{twc98:9b, twc99:6}.
Therefore, the general solution (\ref{eq:gwr_sol}) of the GW relation
can be regarded as a topologically invariant transformation
( i.e., $ \mbox{index}(D) = \mbox{index}(D_c) $ ) which
can transform a nonlocal $ D_c $ into a local $ D $.
Conversely, the transformation (\ref{eq:gwr_sol_Dc}) can transform a
local $ D $ into the non-local $ D_c $.
For a given $ D_c $, the set of transformations,
\{~$ \CT(R) : D=D_c( \Id + R D_c )^{-1} $~\},
form an abelian group with parameter space \{~$R$~\} \cite{twc99:6}.
In general, for any lattice Dirac
operator $ D $ ( not necessarily satisfying the GW relation ),
we can use the topologically invariant transformation
$ D'= D( \Id + R D )^{-1} $ to manipulate its locality.

The next question is whether we can gain anything
( e.g., improving the locality of $ D $ ) by using another
functional form of $ R $ rather than the simplest choice $ R = r \Id $.
We investigate this question by numerical experiments.
For simplicity, we consider the Neuberger-Dirac operator \cite{hn97:7}
\bea
\label{eq:Dh}
D_h = \Id + V, \hspace{4mm}
V = D_w ( D_w^{\dagger} D_w )^{-1/2}
\eea
where $ D_w $ is the Wilson-Dirac fermion operator with negative mass
$ -1 $
\bea
D_w = - 1
      + \frac{1}{2} [ \gamma_{\mu} ( \nabstar{\mu} + \nab{\mu} ) -
                      \nabstar{\mu} \nab{\mu} ]
\eea
where $ \nab{\mu} $ and $ \nabstar{\mu} $ are the forward and
backward difference operators defined in the following,
\bea
\nab{\mu}\psi(x) &=& \
 U_\mu(x)\psi(x+\hat{\mu})-\psi(x)  \nonumber \\
\nabstar{\mu} \psi(x) &=& \ \psi(x) -
 U_\mu^{\dagger}(x-\hat{\mu}) \psi(x-\hat{\mu}) \ .  \nonumber
\eea
The Neuberger-Dirac operator $ D_h $ satisfies the GW relation (\ref{eq:gwr})
with $ R = 1/2 $. Then Eq. (\ref{eq:gwr_sol_Dc}) gives
\bea
D_c = 2 \frac{ \Id + V }{ \Id - V } \ .
\label{eq:DcV}
\eea
Substituting (\ref{eq:DcV}) into the general solution (\ref{eq:gwr_sol}),
we obtain
\bea
\label{eq:ND}
D = 2 ( \Id + V )[ ( \Id - V ) + 2 R ( \Id + V ) ]^{-1} \ .
\eea
For a fixed gauge background, we investigate the locality of
$ D(x,y) $ versus the functional form of $ R(x,y) $. For simplicity,
we consider $ D $ in a two dimensional $ U(1) $ background gauge field
with non-zero topological charge, and we use the same notations for
the background gauge field as Eqs. (7)-(11) in ref. \cite{twc98:4}.
Although it is impossible for us to go through all different functional
forms of $ R(x,y) $, we can use the exponential function
\bea
\label{eq:Rxy}
R(x,y) = r \ \exp( - m | x - y | )
\eea
as a prototype to approximate other forms by varying the parameters
$ r $ and $ m $. In the limit $ m \to 0 $, $ R(x,y) $ is nonlocal,
while in the limit $ m \to \infty $, $ R \to r \Id $ which is the most
ultralocal. Hence, by varying the value of $ m $ from $ 0 $ to
$ \alpha \gg 1 $,
we can cover a wide range of $ R(x,y) $ of very different behaviors.

One of the physical quantities which are sensitive to the locality
of $ D $ is the anomaly function
\BA
\label{eq:ax_gw}
\anx = \tr \left[ \gamma_5 ( R D ) (x,x)
                   +\gamma_5 ( D R ) (x,x)  \right]
\EA
which can serve as an indicator of the localness of $ D $. Since the
Neuberger-Dirac operator is {\it topologically proper} for smooth gauge
backgrounds, the index of $ D $ in (\ref{eq:ND}) is equal to the
background topological charge $ Q $,
\bea
\label{eq:index_thm_a}
\mbox{index}(D) = n_{-} - n_{+} = Q \ .
\eea
This implies that the sum of the anomaly function over all sites on a
finite lattice must be equal to two times of the topological charge
\cite{twc99:6}
\beq
\label{eq:index_thm_b}
  \sum_{x} \anx = 2 ( n_{-} - n_{+} ) = 2 Q =
  \left\{  \begin{array}{ll}
  \frac{1}{16 \pi^2} \sum_x \epsilon_{\mu\nu\lambda\sigma} \
                          F_{\mu\nu}(x) \ F_{\lambda\sigma}(x) \ ,
                          & \mbox{ d = 4 } \ ;       \\
                          \\
  \frac{1}{2 \pi} \sum_x \epsilon_{\mu\nu} \
                          F_{\mu\nu}(x) \ ,
                          & \mbox{ d = 2 } \ .   \\
           \end{array}
           \right.
\eeq
This is true for any $ R $ since the index of $ D $ is invariant under
the transformation (\ref{eq:gwr_sol}),
i.e., $ \mbox{index}(D) = \mbox{index}(D_c) $.
First we consider the gauge configuration with constant field tensors.
If $ D $ is local, then we can deduce that $ \anx $ is constant
for all $x$. From (\ref{eq:index_thm_b}), it follows that
\beq
\label{eq:anx_rho}
\anx = \rho(x) \equiv
  \left\{  \begin{array}{ll}
           \frac{1}{16 \pi^2} \ \epsilon_{\mu\nu\lambda\sigma} \
                          F_{\mu\nu}(x) \ F_{\lambda\sigma}(x) \ ,
                          & \mbox{ d = 4 } \ ;        \\
                          \\
           \frac{1}{2 \pi} \ \epsilon_{\mu\nu} \
                          F_{\mu\nu}(x) \ ,
                          & \mbox{ d = 2 } \ ,        \\
           \end{array}
           \right.
\eeq
where $ \rho(x) $ is the Chern-Pontryagin density in continuum.
Note that Eq. (\ref{eq:anx_rho}) also implies that $ \anx $ is
independent of $ R $ if $ D $ is local.
Next we introduce local fluctuations to the constant background gauge
field, with the topological charge fixed.
Then we expect that (\ref{eq:anx_rho}) remains valid provided
that the locality of $ D $ is not destroyed by the roughness of the
gauge field. Therefore, in general, by comparing the anomaly function
$ \anx $ at each site with the Chern-Pontryagin density $ \rho(x) $,
in a prescribed background gauge field, we can reveal whether
$ D $ is local or not in this gauge background. This provides
another scheme to examine the locality of $ D $ rather than checking
how well $ | D(x,y) | $ can be fitted by an exponentially decay
function. We will use both methods in our investigations.
We define the deviation of the anomaly function
as
\bea
\label{eq:delta}
\delta_D \equiv \frac{1}{N_s} \sum_{x} \frac{|\anx - \rho(x)|}{|\rho(x)|}
\eea
where the summation runs over all sites, and $ N_s $ is the total number
of sites.

{\footnotesize
\begin{table}
\begin{center}
\begin{tabular}{|c|c|c|c|}
\hline
 r  &  m  &  $ \delta_D $ & \mbox{index}(D) \\
\hline
\hline
 0.5  &  0.5  &   1.366  &  1 \\
\hline
 0.5  &  1.0  &   0.5360 &  1 \\
\hline
 0.5  &  2.0  &   0.2086 &  1 \\
\hline
 0.5  &  5.0  &   $ 7.209 \times 10^{-3} $ &  1 \\
\hline
 0.1  &  5.0  &   $ 6.554 \times 10^{-2} $ &  1 \\
\hline
 0.5  &  $ \gg 1 $ &  $ 3.848 \times 10^{-4} $ & 1 \\
\hline
 1.0  &  $ \gg 1 $ &  $ 1.618 \times 10^{-4} $ & 1 \\
\hline
\end{tabular}
\end{center}
\caption{ The deviation of the chiral anomaly function
$ \delta_D $ [ Eq. (\ref{eq:delta}) ] versus
$ R(x,y) $ with parameters $ r $ and $ m $ defined in Eq. (\ref{eq:Rxy}),
on a $ 12 \times 12 $ lattice,
in a constant background gauge field with topological charge $ Q = 1 $.
}
\label{table:anx}
\end{table}
}

In Table \ref{table:anx}, we list the deviation of the chiral anomaly
function, $ \delta_D $, versus $ R(x,y) $ with parameters $ r $ and $ m $
defined in Eq. (\ref{eq:Rxy}), on a $ 12 \times 12 $ lattice,
in a constant background gauge field with topological charge $ Q = 1 $.
The last two rows with $ m \gg 1 $ corresponds to $ R = r \Id $,
and they have the smallest deviations.
They are both local, which can be checked explicitly by plotting
$ | D(x,y) | $  versus $ | x - y | $, as shown in Fig. 1.
For $ r = 1/2 $ ( the row on top of
the last row ), it corresponds to the Neuberger-Dirac operator.
In the first row, $ R(x,y) $ is nonlocal since $ m = 1/2 < 1 $.
It produces a nonlocal $ D $, as shown in Fig. 2, so the resulting
chiral anomaly is very different from the Chern-Pontryagin density
and thus $ \delta_D $ is very large.
Now we increase $ m $ to $ 1 $, $ 2 $ and $ 5 $ successively,
then $ R $ and $ D $ both become more and more local,
as shown in Fig. 3, thus $ \delta_D $ becomes smaller and smaller,
as shown in the second, third and fourth rows of Table 1.
These results indicate that
{\it a nonlocal $ R $ does not produce a local $ D $}, and
{\it a local $ R $ does not make $ D $ more local than that
using $ R = r \Id $}.
For $ m = 5.0 $, if we decrease $ r $ to $ 0.1 $, then $ D $ becomes
very nonlocal, as shown in Fig. 4. Consequently, its $ \delta_D $
( in the fifth row ) is about 10 times larger than that of $ r = 0.5 $
( in the fourth row ).
This suggests that {\it on a finite lattice, $ r $
cannot be too small, otherwise $ D $ will become nonlocal.}

\psfigure 5.0in -0.2in {fig:neu_10} {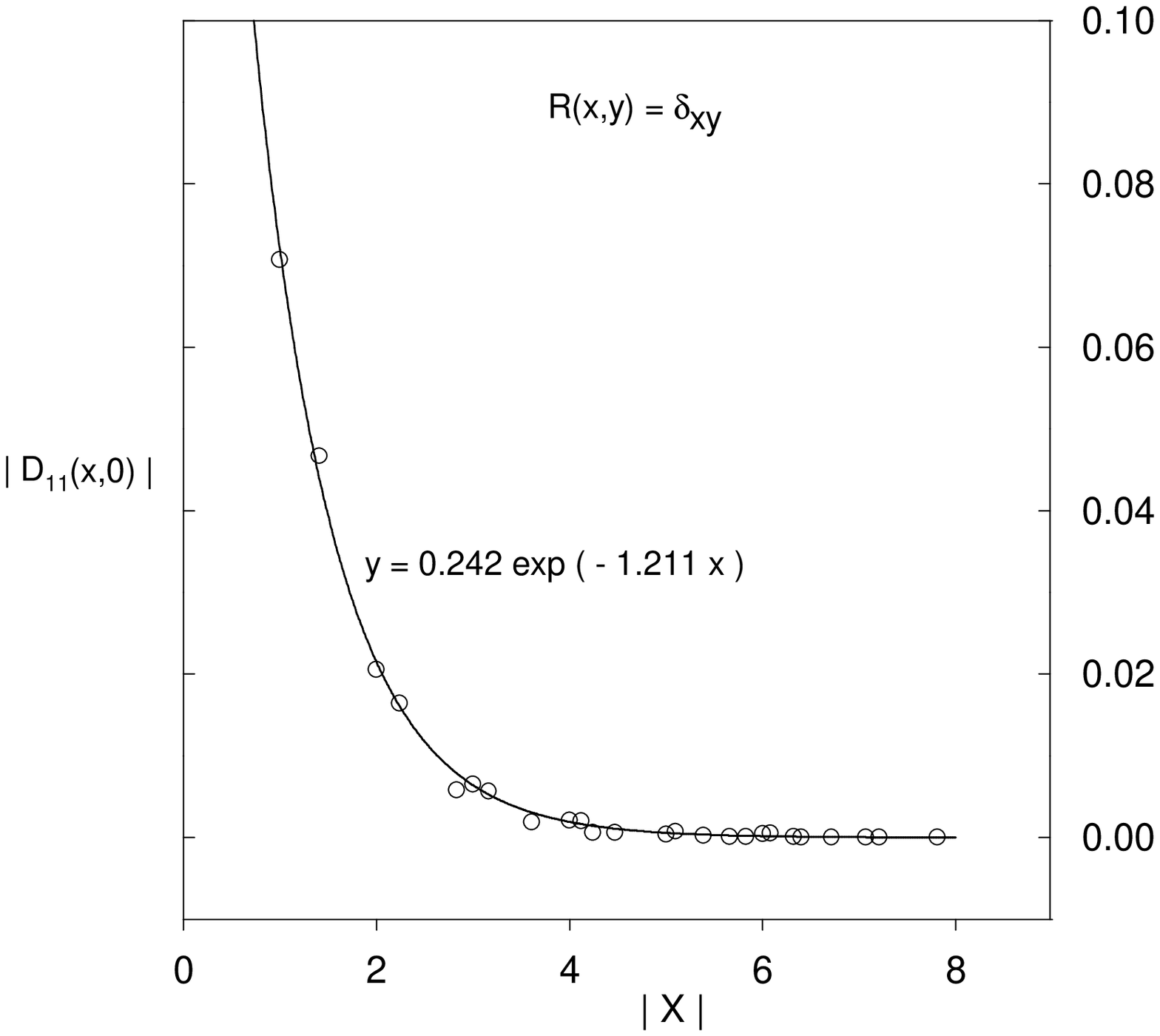} {
One of the Dirac components of $ D(x,0) $,
$ | D_{11} (x,0) | $, is plotted as a function of $ | x | $
for $ R(x,y) = \delta_{xy} $.
The lattice is $ 12 \times 12 $ with periodic
boundary conditions. The constant background gauge field has
topological charge $ Q = 1 $. All data points at the
same distance $ | x | $ from the origin have been averaged.
The solid line is an exponential fit to the data points.
The same decay constant also fits very well for all other Dirac components
of $ D(x,y) $ and for any reference point $ y $. }

\psfigure 5.0in -0.2in {fig:0505} {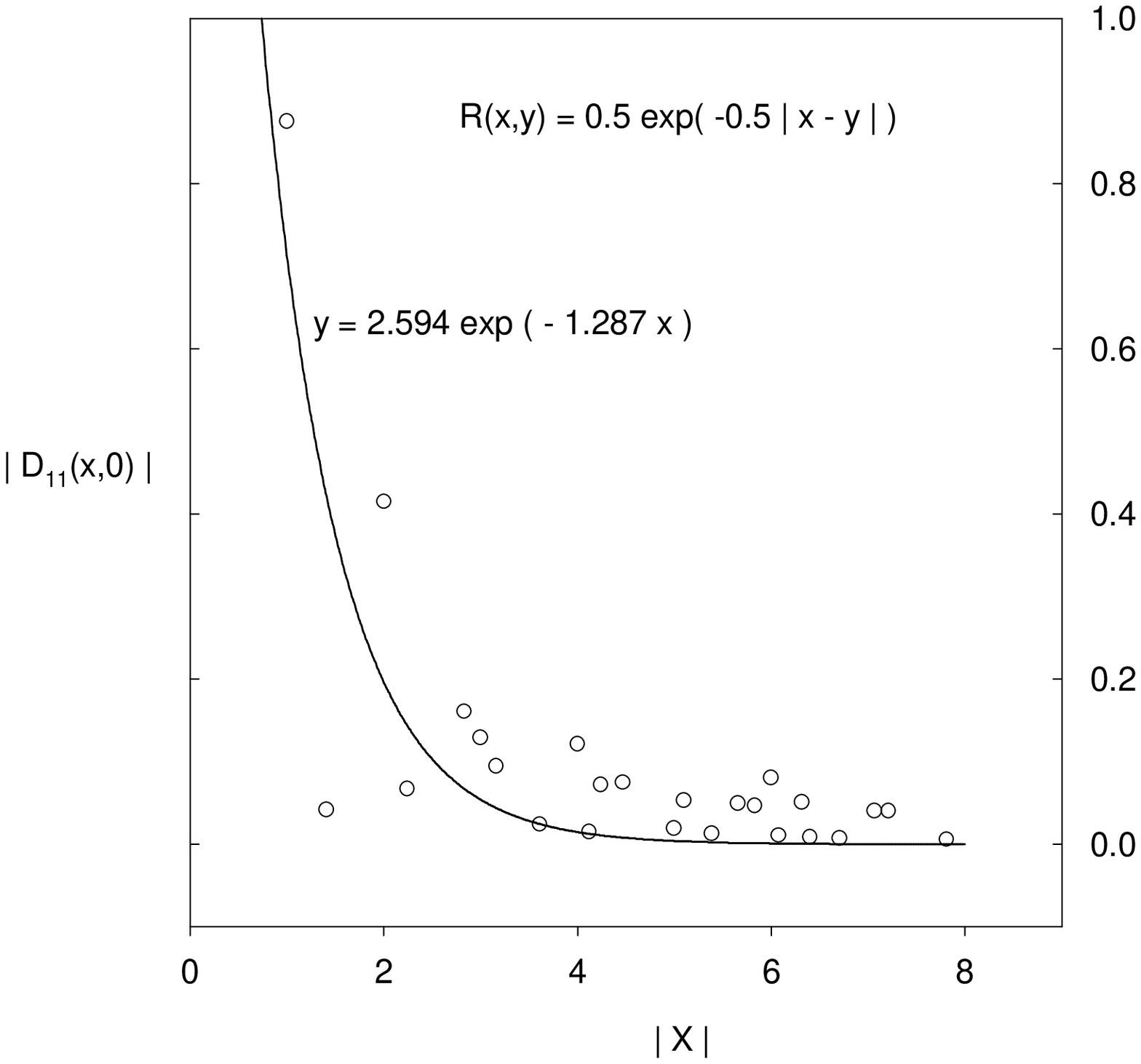} {
One of the Dirac components of $ D(x,0) $ is plotted
as a function of $ | x | $ with $ R(x,y) = 0.5 \exp( -0.5 |x-y| ) $.
Other descriptions are the same as Fig. 1. The non-localness of $ D $
is shown clearly. }

\psfigure 5.0in -0.2in {fig:0550} {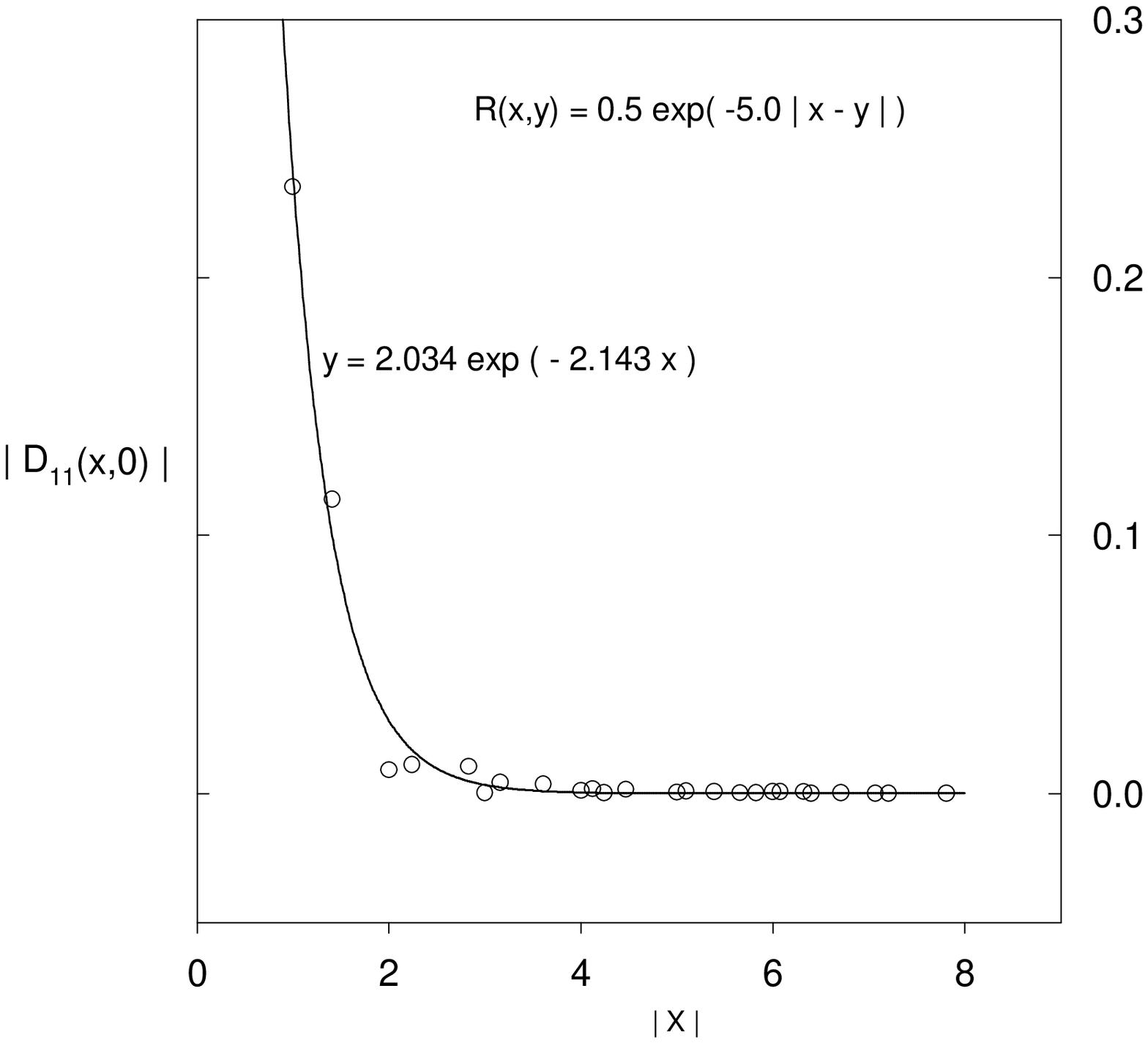} {
One of the Dirac components of $ D(x,0) $ is plotted
as a function of $ | x | $ with $ R(x,y) = 0.5 \exp( -5.0 |x-y| ) $.
Other descriptions are the same as Fig. 1. }

\psfigure 5.0in -0.2in {fig:0150} {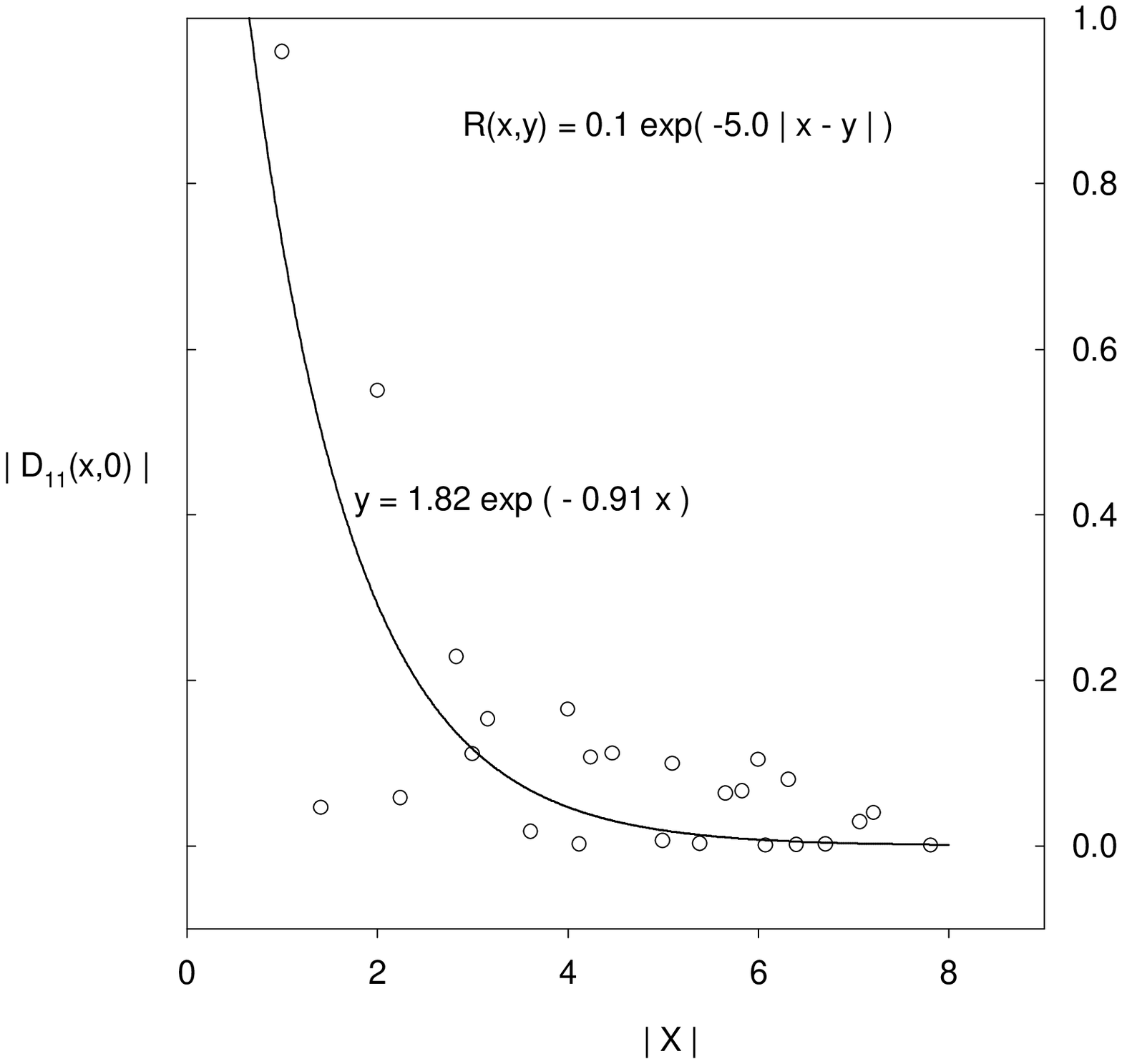} {
One of the Dirac components of $ D(x,0) $ is plotted
as a function of $ | x | $ with $ R(x,y) = 0.1 \exp( -5.0 |x-y| ) $.
Other descriptions are the same as Fig. 1. The non-localness of $ D $
is shown clearly.}

Our numerical results listed in Table 1 as well as those plotted in
Figs. 1-4 strongly suggest that we {\it do not gain anything by
using other functional forms of $ R(x,y) $ than the simplest
choice $ R(x,y) = r \ \delta_{x,y} $}. However, the value of $ r $
plays the important role in determining the localness of $ D $.
We have also tested other functional forms of $ R(x,y) $ as well as
many different gauge configurations. The results from all
these studies are consistent with the conclusion
that {\it the optimal choice for $ R $ is $ R(x,y) = r \ \delta_{x,y} $}.

Now we come to the question concerning the range of proper values of $ r $.
We have already known that $ r $ cannot be zero or very small, otherwise
$ D $ is nonlocal. On the other hand, $ r $ cannot be too large,
otherwise $ D $ is highly peaked in the diagonal elements
( i.e., $ D_{\alpha\beta}(x,y) \sim
D_{\alpha\alpha}(x,x) \delta_{\alpha\beta} \delta_{x,y} $ ), which is
unphysical since it does not respond properly to the background gauge field
( e.g., the chiral anomaly is incorrect even though the index of $ D $
is equal to the background topological charge ).
In Table \ref{table:anx_rr}, we list the deviation of
the chiral anomaly function, $ \delta_D $, versus $ R(x,y) = r \ \delta_{x,y} $,
on a $ 12 \times 12 $ lattice ( the second coluum ),
in a constant background gauge field with topological charge $ Q = 1 $.
We see that the proper values of $ r $ are approximately in the
range $ 0.5 \sim 1.2 $, where $ D $ can reproduce the
continuum chiral anomaly precisely. Next we investigate how the lattice
size affects the range of proper values of $ r $.
The results of $ \delta_D $ for lattice sizes $ 16 \times 16 $
and $ 20 \times 20 $ are listed in the third and the fourth coluums in
Table \ref{table:anx_rr}.
They clearly show that the lower bound of $ r $ can be pushed to a smaller
value, $ \sim 0.2 $, when the size of the lattice is increased to
$ 20 \times 20 $.
Therefore, it suggests that the chiral limit (~$ r \to 0 $~and~$ D \to D_c $~)
can be approached by decreasing the value of $ r $ while increasing the
size of the lattice, at finite lattice spacing. This provides a
{\it nonperturbative definition of the chiral limit} for any $ D $ of the
general solution (\ref{eq:gwr_sol}) with $ D_c $ satisfying the
necessary physical requirements \cite{twc98:9a}.

It is evident that the range of proper values of $ r $ also depends on the
background gauge configuration.
However, we suspect that when the background gauge configuration becomes
very rough, there may not exist any values of $ r $ such that the chiral
anomaly function is in good agreement with the Chern-Pontryagin density.
We intend to return to this question in a later publication.

{\footnotesize
\begin{table}
\begin{center}
\begin{tabular}{|c|c|c|c|c|}
\hline
 r  &  $ \delta_D $  &  $ \delta_D $   &  $ \delta_D $   & \mbox{index}(D) \\
    & ( $ 12 \times 12 $ ) & ( $ 16 \times 16 $ ) & ( $ 20 \times 20 $ ) & \\
\hline
\hline
 0.1  & $ 6.232 \times 10^{-2} $  & $ 1.792 \times 10^{-2} $
      & $ 4.854 \times 10^{-3} $  &  1 \\
\hline
 0.2  & $ 4.818 \times 10^{-3} $  & $ 6.229 \times 10^{-4} $
      & $ 7.750 \times 10^{-5} $  &  1 \\
\hline
 0.5  & $ 3.848 \times 10^{-4} $  & $ 2.434 \times 10^{-5} $
      & $ 1.559 \times 10^{-6} $  &  1 \\
\hline
 0.8  & $ 1.698 \times 10^{-4} $  & $ 8.268 \times 10^{-6} $
      & $ 4.976 \times 10^{-7} $  &  1 \\
\hline
 1.0  & $ 1.618 \times 10^{-4} $  & $ 6.596 \times 10^{-6} $
      & $ 3.432 \times 10^{-7} $  &  1 \\
\hline
 1.2  & $ 3.448 \times 10^{-4} $  & $ 1.318 \times 10^{-5} $
      & $ 5.308 \times 10^{-7} $  &  1 \\
\hline
 1.5  & $ 1.688 \times 10^{-3} $  & $ 1.243 \times 10^{-4} $
      & $ 8.783 \times 10^{-6} $  &  1 \\
\hline
 2.0  & $ 9.940 \times 10^{-3} $  & $ 1.569 \times 10^{-3} $
      & $ 2.348 \times 10^{-4} $  &  1 \\
\hline
 5.0  & $ 0.1599 $ &  $ 9.314 \times 10^{-2} $
      & $ 5.088 \times 10^{-2} $ & 1 \\
\hline
\end{tabular}
\end{center}
\caption{ The deviation of the chiral anomaly function,
$ \delta_D $ [ Eq. (\ref{eq:delta}) ], versus
$ R(x,y) = r \ \delta_{x,y} $, for lattice sizes, $ 12 \times 12 $,
$ 16 \times 16 $ and $ 20 \times 20 $ respectively.
The background gauge field has constant field tensors
with topological charge $ Q = 1 $.
The index of $ D $ is always equal to one in each case. }
\label{table:anx_rr}
\end{table}
}

In summary, we have clarified the role of $ R $ in the general
solution (\ref{eq:gwr_sol}) of the Ginsparg-Wilson relation.
It provides a topologically invariant transformation which transforms
the chirally symmetric and nonlocal $ D_c $ into a local $ D $ which
satisfies the GW relation, the exact chiral symmetry on the lattice.
Having $ R $ local in the position space is a necessary
condition to ensure the absence of additive mass renormalization in
the fermion propagator, as well as to produce a local $ D $,
which is vital for obtaining the correct chiral anomaly.
Our numerical results strongly
suggest that the optimal form of $ R $ is $R(x,y) = r \ \delta_{x,y}$.
The range of proper values of $ r $ depends on the background gauge
configuration as well as the size of the lattice, $ L = N a $.
In the limit $ N \to \infty $, for smooth gauge backgrounds,
the lower bound of proper values of $ r $
goes to zero, thus the chiral limit ( $ r \to 0 $ and $ D \to D_c $ )
can be approached nonperturbatively at finite lattice spacing.

\bigskip
\bigskip


\flushpar
{\bf Acknowledgement }
\medskip

\noindent
I would like to thank all participants of Chiral '99, workshop on
chiral gauge theories ( Taipei, Sep. 13-18, 1999 ), for their
stimulating questions and interesting discussions.
I am also indebted to Herbert Neuberger for his helpful comments
on the first version of this paper.
This work was supported by the National Science Council,
R.O.C. under the grant number NSC89-2112-M002-017.

\bigskip
\bigskip

\vfill\eject

\vfill\eject

\end{document}